\documentclass{ws-mpla}
\usepackage{amsfonts}
\usepackage{amsmath}
\usepackage{amssymb}
\usepackage{graphicx}
\usepackage{hyperref}

\begin{document}

\title{Induced Phantom and 5D Attractor Solution in Space-Time-Matter Theory}
\author{Hongya Liu\thanks{
Corresponding author: hyliu@dlut.edu.cn}, Huanying Liu, Baorong Chang, Lixin
Xu}

\address{Department of Physics, Dalian University of Technology,
Dalian,  Liaoning, 116024, P.R. China}

\maketitle

\pub{Received (Day Month Year)}{Revised (Day Month Year)}

\begin{abstract}
In Spacetime-Matter theory we assume that the $4D$ induced matter
of the $5D$ Ricci-flat bouncing cosmological solutions contains a
perfect fluid as well as an induced scalar field. Then we show
that the conventional $4D$ quintessence and phantom models of dark
energy could be recovered from the $5D$ cosmological solutions. By
using the phase-plane analysis to study the stability of evolution
of the $5D$ models, we find that the conventional $4D$ late-time
attractor solution is also recovered. This attractor solution
shows that the scale factors of the phantom dominated universes in
both the $4D$ and $5D$ theories will reach infinity in a finite
time and the universes will be ended at a new kind of spacetime
singularity at which everything will be annihilated. We also find
that the repulsive force of the phantom may provide us with a
mechanics to explain the bounce.
\end{abstract}


\ccode{PACS Nos.: 04.50.+h, 98.80.-k.}

\section{Introduction}

In recent decades the combinations of observations from high
redshift supernovae, Cosmic Microwave Background Radiation(CMBR)
\cite{Bernardis} and the galaxy redshift surveys indicate the
existence of an exotic component with negative pressure dubbed
dark energy. This dark energy violates strong
energy condition and drives the universe accelerating at present time. \cite%
{Riess}\cite{Tonry}\cite{A.G.Riess} Dark energy and accelerating
universe have been discussed extensively from different points of
view. A natural candidate for this exotic component is the
cosmological constant for which, however, there exists serious
problems such as the fine tuning problem and the coincidence
problem. So, inspired by inflation, a scalar field which is
minimally coupled to the conventional matter has been treated as a
dynamic dark energy model, called quintessence. In the
quintessence model \cite{Zlatev} some special scalar potentials
are designed to obtain track or attractor solutions in which the
initial information is eliminated in the late time of the universe
and so the fine tuning problem is resolved. However, in
quintessence dark energy model, the equation of state is in the
range $[-1,1]$ which does not meet the observations well; The
observations favor a value which is less than $-1$. So another
dynamic dark energy model, called phantom, \cite{Caldwell} was
presented, though there exists a `Big Rip' \cite{S. N} problem in
the phantom model. Also, another consistent (without ghosts,
phantoms and Big Rip) Dark Energy model for the Super-Accelerated
phase $(w<-1)$ of the cosmic expansion are presented in \cite{OW}.

In Kaluza-Klein theories as well as in brane world scenarios, our $4D$
universe is believed to be embedded in a higher-dimensional manifold. Here,
in this paper, we consider Wesson's Spacetime-Matter (STM) theory \cite%
{Wesson}\cite{Overduin}. This theory is distinguished from the
classical Kaluza-Klein theory by that it has an noncompact fifth
dimension and that it is empty viewed from $5D$ and sourceful
viewed from $4D$. Because of this, the STM theory is also called
induced matter theory and the effective $4D$ matter is also called
induced matter. That is, in STM theory, the $5D$ manifold is
Ricci-flat while the $4D$ hypersurface is curved by the $4D$
induced matter. Mathematically, this approach is supported by
Campbell's theorem that any analytical solution of the
$N$-dimensional Einstein equations with a source can be locally
embedded in an $(N+1)$-dimensional Ricci-flat manifold
\cite{Campbell}.

A class of $5D$ cosmological solutions of the STM theory was
originally given by Liu and Mashhoon \cite{LiuB} and restudied
later by Liu and Wesson \cite{Liu}. This class of exact solutions
satisfy the $5D$ Ricci-flat equations $R_{AB}=0$ and is
algebraically rich because it contains two arbitrary functions of
the time $t$. It was shown \cite{Liu} that several properties
characterize these $5D$ models: Firstly, the $4D$ induced matter
could be described by a perfect fluid plus a variable cosmological
`constant'. Secondly, by properly choosing the two arbitrary functions of $t$%
, both the matter-dominated and radiation-dominated standard FRW models
could be recovered. In the third, the big bang singularity of the $4D$
standard cosmology is replaced by a big bounce at which the size of the
universe is finite, and before the bounce the universe contracts and after
the bounce the universe expands. Many further studies could be found in
literature such as those about the embeddings of the $5D$ bounce solutions
to brane models \cite{Seahra}, about the acceleration of the $5D$ universe
\cite{LiuXu}, and about the isometry between the $5D$ solutions and a kind
of $5D$ topological black holes \cite{SearhaW}.

In this paper we are going to introduce a scalar field in the $5D$
bounce cosmological solutions and build a $5D$ quintessence and
phantom model for dark energy. Rather than assuming the scalar
field $\phi$ being the
Kaluza-Klein dilation for which $\phi$ is related to the metric component $%
g_{55}$ in an explicit form $g_{55}\propto \phi ^{2}$, we will
just assume that the induced matter being composed of two
components: the conventional perfect fluid and a scalar field. We
will show that the $4D$ conventional quintessence and phantom
models as well as their late time attractor solutions could be
recovered in the bounce solutions.

\section{Induced Scalar Field in the $5D$ Cosmological models}

The $5D$ metric of the bounce solutions reads \cite{LiuB}
\begin{equation}
dS^{2}=B^{2}dt^{2}-A^{2}(\frac{dr^{2}}{1-kr^{2}}+r^{2}d\Omega ^{2})-dy^{2},
\label{1}
\end{equation}
where, $d\Omega ^{2}\equiv (d\theta ^{2}+\sin ^{2}\theta d\phi ^{2})$ and
\begin{eqnarray}
A^{2} &=&(\mu ^{2}+k)y^{2}+2\nu y+\frac{\nu ^{2}+K}{\mu ^{2}+k},  \label{2}
\\
B &=&\frac{1}{\mu }\frac{\partial A}{\partial t}\equiv \frac{\dot{A}}{\mu}.
\notag
\end{eqnarray}
Here $\mu =\mu (t)$ and $\nu =\nu(t)$ are two arbitrary functions of $t$, $k$
is the $3D$ curvature index, $k=\pm 1,0$, and $K$ is a constant. This class
of solution satisfies the $5D$ vacuum equations $R_{AB}=0$. Thus we have $%
I_{1}\equiv R=0$ and $I_{2}\equiv R^{AB}R_{AB}=0$. The third invariant is
found to be
\begin{equation}
I_{3}=R_{ABCD}R^{ABCD}=\frac{72K^{2}}{A^{8}},  \label{3}
\end{equation}
which shows that $K$ determines the curvature of the $5D$ manifold. So the $%
5D$ manifold is curved if $K\neq 0$.

The $4D$ metric contained in the $5D$ line element (\ref{1}) is
\begin{equation}
ds^{2}=g_{\mu \nu }dx^{\mu }dx^{\nu }=B^{2}dt^{2}-A^{2}(\frac{dr^{2}}{%
1-kr^{2}}+r^{2}d\Omega ^{2}).
\end{equation}
One can use this $4D$ metric to calculate the nonvanishing
components of the $4D$ Einstein tensor, then one obtains
\begin{eqnarray}
^{(4)}G_{0}^{0} &=&\frac{3(\mu ^{2}+\kappa )}{A^{2}}  \label{4},
\\
^{(4)}G_{1}^{1} &=&^{(4)}G_{2}^{2}=^{(4)}G_{3}^{3}=\frac{2\mu \dot{\mu}}{A%
\dot{A}}+\frac{\mu ^{2}+\kappa }{A^{2}}.  \label{5}
\end{eqnarray}

Generally speaking, the $4D$ Einstein tensor in (\ref{4}) and
(\ref{5}) can define a $4D$ effective or induced energy-momentum
tensor $^{\left( 4\right) }T_{\mu }^{\nu }$ via $^{\left( 4\right)
}G_{\mu }^{\nu }=\chi ^{2(4)}T_{\mu }^{\nu }$ with $\chi ^{2}=8\pi
G$ as is done in the STM theory. Note that this kind of definition
for $^{\left( 4\right) }T_{\mu }^{\nu }$ is not completely
arbitrary; the R.H.S. of (\ref{4}) and (\ref{5}) play the role as
constraints on the forms of $^{\left( 4\right) }T_{\mu }^{\nu }$.
In Ref. 12, this energy-momentum tensor was assumed to be a
perfect fluid
with density $\rho $ and pressure $p$ plus a variable cosmological term $%
\Lambda $, and the result shows that this assumption works well. Surely,
this\ $\Lambda $ could be served as a candidate for dark energy. However,
nowadays more hopeful candidates for dark energy seem to be the scalar field
models such as the quintessence and phantom. So, in this paper, we assume
the $4D$ induced energy-momentum tensor consisting of two parts:
\begin{align}
^{\left( 4\right) }T_{\mu \nu }& =T_{\mu \nu }^{m}+T_{\mu \nu }^{\phi },
\notag \\
T_{\mu \nu }^{m}& =(\rho _{m}+p_{m})u_{\mu }u_{\nu }-p_{m}g_{\mu \nu },
\notag \\
T_{\mu \nu }^{\phi }& =\varepsilon \partial _{\mu }\phi \partial _{\nu }\phi
-g_{\mu \nu }\left[ \frac{1}{2}\varepsilon g^{\mu \nu }\partial _{\mu }\phi
\partial _{\nu }\phi -V\left( \phi \right) \right] ,  \label{6}
\end{align}%
where $T_{\mu \nu }^{m}$ is a perfect fluid energy-momentum tensor and $%
T_{\mu \nu }^{\phi }$ a scalar field one with $\varepsilon =\pm 1$. For $%
\varepsilon =+1$, $T_{\mu \nu }^{\phi }$\ is regular and it represents the
quintessence model. For $\varepsilon =-1$, the scalar field has a negative
dynamic energy and $T_{\mu \nu }^{\phi }$ represents the phantom model. We
will show in the following that this kind of assumption in (\ref{6}) will
work well.

Furthermore, we assume that each of the two components $T_{\mu \nu }^{m}$
and $T_{\mu \nu }^{\phi }$ conserves independently just as in most of the $4D
$ dark energy models. That is, we assume both the $4D$ conservation laws $%
T_{\mu ;\nu }^{\left( m\right) \nu }=0$ and $T_{\mu ;\nu }^{\left( \phi
\right) \nu }=0$ hold. We should also notice that the coordinate time $t$ is
not the proper time in the solutions (\ref{1}) and (\ref{2}), and,
generally, one can not transform it to the proper time without changing the
form of the $5D$ metric (\ref{1}).\ However, on a given hypersurface $y=$%
const., the proper time $\tau $ relates the coordinate time $t$
via $d\tau =Bdt$. So the proper definitions for the Hubble
parameter $H$ and the deceleration parameter $q$ can be given as
follows
\begin{equation}
H\equiv \frac{\dot{A}}{AB}=\frac{\mu }{A},\quad q=-\frac{A\dot{\mu}}{\mu
\dot{A}},  \label{H,q}
\end{equation}
where the second equation in (\ref{2}) was used. Using these relations in (%
\ref{4})-(\ref{6}) and in the conservation law $T_{\mu ;\nu }^{\left(
m\right) \nu }=0$ and $T_{\mu ;\nu }^{\left( \phi \right) \nu }=0$, one
obtain the $4D$ Einstein equations and the equations of motion for the
scalar field being:
\begin{eqnarray}
\dot{\rho}_{m}+3\frac{\dot{A}}{A}\left( \rho _{m}+p_{m}\right) =0,  \notag \\
\ddot{\phi}+\left( 3\frac{\dot{A}}{A}-\frac{\dot{B}}{B}\right) \dot{\phi}%
+\varepsilon B^{2}\frac{dV\left( \phi \right) }{d\phi } =0,  \notag \\
H^{2}+\frac{k}{A^{2}} =\frac{k^{2}}{3}(\rho _{m}+\rho _{\phi }),  \notag \\
\dot{H} =-\frac{k^{2}B}{2}(\rho _{m}+p_{m}+\rho _{\phi }+p_{\phi
}) \label{7},
\end{eqnarray}%
where
\begin{equation}
\rho _{\phi }\equiv \frac{1}{2}\varepsilon \frac{\dot{\phi}^{2}}{B^{2}}%
+V\left( \phi \right) ,\quad p_{\phi }\equiv \frac{1}{2}\varepsilon \frac{%
\dot{\phi}^{2}}{B^{2}}-V\left( \phi \right)  \label{8}
\end{equation}%
are the energy density and pressure of the $\phi $ field, respectively, and
the potential $V(\phi )$ is assumed to be exponentially dependent on $\phi $
by $V\left( \phi \right) =V_{0}\exp \left( -\lambda k\phi \right) $. The
equation-of-state parameter for the scalar field is found to be
\begin{equation}
\omega _{\phi }=\frac{p_{\phi }}{\rho _{\phi }}=\frac{\varepsilon \dot{\phi}%
^{2}-2B^{2}V\left( \phi \right) }{\varepsilon \dot{\phi}^{2}+2B^{2}V\left(
\phi \right) }.  \label{9}
\end{equation}%
By comparing the results in (\ref{7})-(\ref{9}) with those in the standard $%
4D$ theory, we find that on a given $4D$ hypersurface one has $Bdt=d\tau $
and the two theories turn out to be of completely the same form. So we
conclude that the $4D$ quintessence and phantom models of dark energy are
recovered in the $5D$ big bounce model. We should also point out that the
Hubble parameter $H$ in (\ref{7}) is constrained by the relation in (\ref%
{H,q}), $H=\mu /A$. So the global evolution of the scale factor $A$ may
differ from that in the $4D$ conventional theory as we will see in the next
section.

\section{5D Attractor Solution of Phantom Model}

The quintessence model with $\varepsilon =+1$ was discussed in a
previous work \cite{Chang}. Here, in this section, we will focus
on the $\varepsilon =-1$ case, i.e. the phantom model of the dark
energy.

Similarly as in Ref. 17, for $\varepsilon =-1$ we define $x$ and $%
y$ in a plane-autonomous system as
\begin{equation}
x\equiv \frac{k\dot{\phi}}{\sqrt{6}BH},\quad y\equiv \frac{k\sqrt{V}}{\sqrt{3%
}H}  \label{10}.
\end{equation}%
For a spatially flat universe($k=0$), we find that eqs. (\ref{7})-(\ref%
{9}) lead to the same evolution equations for $x$ and $y$ as in Ref. \cite%
{Caldwell},
\begin{eqnarray}
x^{\prime } &=&\frac{3}{2}x\left[ -2x^{2}+\gamma
_{m}(1+x^{2}-y^{2})\right] -3x-\sqrt{\frac{3}{2}}\lambda y^{2}
\notag ,\\
y^{\prime } &=&\frac{3}{2}y\left[ -2x^{2}+\gamma _{m}(1+x^{2}-y^{2})\right] -%
\sqrt{\frac{3}{2}}\lambda xy  \label{11},
\end{eqnarray}%
where a prime denotes derivative with respect to\ $N$ with $N=\ln A$. Define
the densities%
\begin{equation}
\Omega _{m}=\frac{k^{2}\rho _{m}}{3H^{2}},\quad \Omega _{\phi }=\frac{%
k^{2}\rho _{\phi }}{3H^{2}}.
\end{equation}%
Then eqs. (\ref{7}) and (\ref{8}) give
\begin{eqnarray}
\Omega _{m}+\Omega _{\phi } &=&1  \notag ,\\
-x^{2}+y^{2} &=&\Omega _{\phi }.
\end{eqnarray}%
Meanwhile, from (\ref{8}) and (\ref{10}), the effective equation of state
for the scalar field is
\begin{equation}
\gamma _{\phi }\equiv \frac{\rho _{\phi }+p_{\phi }}{\rho _{\phi }}=\frac{%
-2x^{2}}{-x^{2}+y^{2}},
\end{equation}%
\begin{equation}
\omega _{\phi }=\frac{p_{\phi }}{\rho _{\phi }}=\frac{x^{2}+y^{2}}{%
x^{2}-y^{2}},
\end{equation}%
where the scalar field is represented by a fluid with a baryotropic equation
of state $p_{\phi }=(\gamma _{\phi }-1)\rho _{\phi }$ and $\gamma _{\phi }$
is a constant with $0\leq \gamma _{\phi }\leq 2$.

As is known from Ref. 6 that there is a stable node in the
phase plane at%
\begin{eqnarray}
x &=&x_{c}=-\frac{\lambda }{\sqrt{6}}\quad ,  \notag \\
y &=&y_{c}=\sqrt{1+\frac{\lambda ^{2}}{6}},  \label{16}
\end{eqnarray}%
at which we have%
\begin{eqnarray}
\omega _{\phi } &=&-\left( 1+\frac{\lambda ^{2}}{3}\right) ,  \notag \\
\Omega _{m} &=&0,\quad \Omega _{\phi }=1.\quad \label{17}
\end{eqnarray}%
The second equation in (\ref{17}) means that the universe is dominated by
the scalar field alone, while the first equation in (\ref{17}) gives $\omega
_{\phi }<-1$ which can better explain the observation for the dark energy.
So the stable node in (\ref{16}) represents an attractor solution which
guides the late time evolution of the universe.

In the following we want to see how the phantom-dominated universe
evolves in the $5D$ bounce model and how much it differs from that
in the $4D$ standard phantom model. To see this, we should firstly
know the scale factor
in the $4D$ one. Note that all the equations from (\ref{H,q}) to (\ref%
{17}) become the same as those in the $4D$ phantom model if $B=1$ and $A=a(t)
$. Then, by solving the last equation in (\ref{7}) we obtain%
\begin{equation}
a(t)=\left( \frac{t_{m}^{2}}{t_{m}-t}\right) ^{2/\lambda ^{2}}\quad ,
\label{18}
\end{equation}%
where $t_{m}$ is a constant. For $t<t_{m}$ the universe expands as $t$
increases. We can easily calculate the decelerating parameter $q$, which is $%
-\left( 1+\lambda ^{2}/2\right) $. As $t\rightarrow t_{m}$, we find $%
a(t)\rightarrow \infty $. Therefore, the $4D$ phantom dominated
universe is expanding and accelerating if $t<t_{m}$, and will
expand to infinity in a finite time $t_{m}$.

Now let us return to the $5D$ exact bounce solutions. Substituting (\ref{16}%
) into (\ref{10}) and using the corresponding relations of $B$ in (\ref{2}), $%
H $ in (\ref{H,q}) and $V\left( \phi \right) =V_{0}\exp \left( -\lambda
k\phi \right) $, we obtain%
\begin{equation}
\frac{d\phi }{d\left( \ln A\right) }=-\frac{\lambda }{k},\quad \quad
V_{0}e^{-\lambda k\phi }=\sqrt{1+\frac{\lambda ^{2}}{6}}\frac{3\mu ^{2}}{%
k^{2}A^{2}}  \label{19}.
\end{equation}%
Thus we obtain%
\begin{equation}
\mu =c_{1}A^{\frac{\lambda ^{2}}{2}+1}  \label{20},
\end{equation}%
where $c_{1}$ is a constant. Note that the scale factor $A$ in the
exact
$5D$ solutions (\ref{1}) and (\ref{2}) depends on two arbitrary functions $%
\mu (t)$ and $\nu (t)$. We do not need $\nu (t)$ in the following.
However, one can in principle get $\nu (t)$ solved out by
substituting (\ref{20}) into (\ref{2}).

Using (\ref{20}) in the expression of $q$ in (\ref{H,q}), we can calculate
the deceleration parameter which is%
\begin{equation}
q=-\frac{A\dot{\mu}}{\mu \dot{A}}=-\left( \frac{\lambda
^{2}}{2}+1\right) \label{21}.
\end{equation}%
Thus we see that the phantom dominated universe in the $5D$ model
is accelerating with $q<-1$. Here we should point out that current
observation of the deceleration parameter is $q_0=-0.67\pm 0.25$
indicating $q_0>-1$. This seemingly contradiction between
observation and the phantom dominated result (\ref{21}) can be
intepreted\ as due to the fact that our present universe also
contains roughly one-third of dark matter plus baryons which drive
the universe to decelerating.

To retain the bounce, we add a new term inside the bracket of (\ref{18}) and
then the scale factor $A$ has the form%
\begin{equation}
A=\left( t_{m}-t+\frac{t_{m}^{2}}{t_{m}-t}\right) ^{2/\lambda ^{2}}.
\label{22}
\end{equation}%
Then (\ref{20}) gives%
\begin{equation}
\mu =c_{1}\left( t_{m}-t+\frac{t_{m}^{2}}{t_{m}-t}\right) ^{\left( 1+\frac{2%
}{\lambda ^{2}}\right) }  \label{23}
\end{equation}%
Here we should mention that (\ref{22}) and (\ref{23}) do satisfy the $5D$
field equations (\ref{7}). We find that if $0<t<t_{m}$, the second term
inside the brackets of (\ref{22}) and (\ref{23}) dominate, so we have%
\begin{equation}
B=\frac{\dot{A}}{\mu }\rightarrow 1\quad \quad \text{as }t\rightarrow t_{m}
\label{24}
\end{equation}%
Therefore we see that when $t\rightarrow t_{m}$ we have $B\rightarrow 1$ and
$t\rightarrow \tau $ and the scale factor $A$ evolves with the same rate as
in the corresponding $4D$ case (\ref{18}). Meanwhile, from (\ref{22}) one
can show that there is a bounce at $t=0$ at which $\dot{A}=0$ and $A$
reaches its minimum $A=\left( 2t_{m}\right) ^{2/\lambda ^{2}}$. The
evolutions of the scale factor $A(t)$ and $a(t)$ with $\lambda =1$ and $%
t_{m}=5$ are plotted in Fig.1.
\begin{figure}[th]
\centerline{\psfig{file=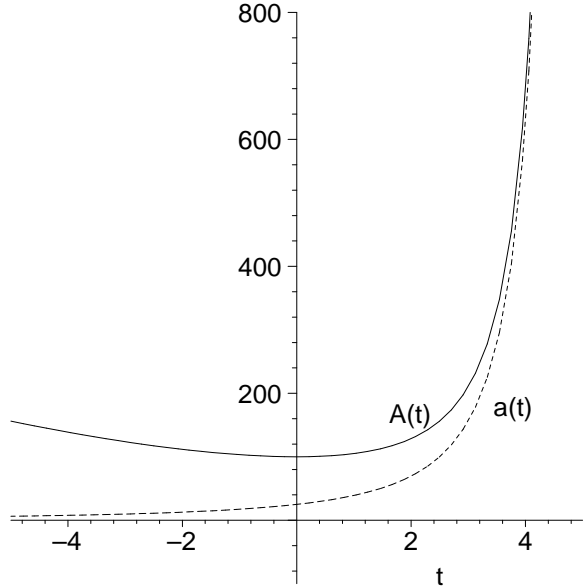,width=3.0in}}
\vspace*{8pt} \caption{The evolutions of the scale factors $a(t)$
(dashed line) and $A(t)$ (solid line) with $\protect\lambda =1$
and $t_{m}=5$ in the phantom
dominated models, where $a(t)$ is the scale factor in the $4D$ model, and $%
A(t)$ is that in the $5D$ bounce model.}
\label{f1}
\end{figure}

\section{Conclusion and Discussion}

The Spacetime-Matter theory is characterized by that the $4D$
matter could be explained as induced from a higher-dimensional
Ricci-flat vacuum. In particular, this induced matter is modelled
by a perfect fluid as well as a cosmological term. However, in
recent years, more and more researchers believe that there is a
$4D$ scalar field which played a very important role in
constructing various dark energy models such as the quintessence
and phantom. In this paper we suppose that this kind of scalar
field is also induced from a higher-dimensional vacuum. That is,
we assume the induced matter being composed of two components: the
usual perfect fluid and a scalar field. Then we find that the
conventional quintessence and phantom models for dark energy could
be recovered from the $5D$ bouncing cosmological solutions.

By using the phase-plane analysis to study the stability of
evolution of this $5D$ cosmological models, we find that the
familiar late-time attractor solution is recovered in the $5D$
theory. As a comparison, we have plotted in Fig. \ref{f1} the two
curves: one is the $4D$ scale factor $a(t)$ (dashed line), another
is the $5D$ one $A(t)$ (solid line). From Fig. \ref{f1} one can
see that both $a(t)$ and $A(t)$ of the phantom-dominated universe
will reach infinity in a finite time and the universe will undergo
a Big Rip and will be ended at a new kind of spacetime singularity
at $t=t_{m}$ at which everything will be annihilated. One also see
in Fig. \ref{f1} that even in the $4D$ theory the time axis can be
traced back to negative infinity. But this would not cause any
problems if one simply assumes that the phantom dominated universe
began at a time after the big bang as required by observation.
However, we should point out that in the $5D$ bouncing model the
curve of $A(t)$ may provide with not only a later-time attractor
solution but also a mechanics to explain the bounce. Generally
speaking, if the universe before and during the bounce was
dominated by phantom, and if matter was created some time after
the bounce, then the repulsive force of the phantom field can
simple explain the bounce.

\section*{Acknowledgments}

This work was supported by NSF (10273004) and NBRP (2003CB716300) of P. R.
China.

\end{document}